\documentclass[useAMS,usenatbib,usegraphicx]{mn2e}
\usepackage{multirow}
\usepackage{url}

\title[On the identification of the \emph{Fermi}/LAT source 0FGL J2001.0+4352 with a BL Lac]{On the 
identification of the \emph{Fermi}/LAT source 0FGL J2001.0+4352 with a BL Lac\thanks{Partly based 
on observations collected at the Astronomical Observatory of Bologna in Loiano
(Italy)}}

\author[L. Bassani et al.]
{L.~Bassani,$^1$\thanks{E-mail address: \texttt{bassani@iasfbo.inaf.it}}, R. Landi,$^{1}$, N. 
Masetti$^{1}$, P. Parisi$^{1}$, A. Bazzano$^{2}$, P. Ubertini,$^{2}$\\
$^1$ INAF/IASF-Bologna, Via P. Gobetti 101, I-40129 Bologna, Italy \\
$^2$ INAF/IASF-Roma, Via Fosso del Cavaliere 100, I-00133, Roma, Italy \\
}

\date{Accepted .... Received ...; in original form ...}
\pagerange{\pageref{firstpage}--\pageref{lastpage}} \pubyear{2008}

\begin{document}
\label{firstpage}
\maketitle 

\begin{abstract}
We report on the identification of the gamma-ray source 0FGL J20001.0+4352 listed in the \emph{Fermi} 
bright
source catalogue. This object, which has an observed 1--100 GeV flux of 
(7.8$\pm$1.2)$\times 10^{-9}$ ph
cm$^{-2}$ s$^{-1}$ and is located close to the Galactic plane, is not associated with any previously 
known
high energy source. We use archival \emph{XMM-Newton} and \emph{Swift}/XRT data to localise with arcsec
accuracy the X-ray counterpart of this GeV emitter and to characterise its X-ray properties:
the source is bright (the 0.2--12 keV flux is $1.9\times10^{-12}$ erg cm$^{-2}$ s$^{-1}$), variable 
(by a factor of $\sim$2) and with 
a steep power law spectrum ($\Gamma=2.7$). It coincides with a radio bright ($\sim$200 mJy at 8.4 GHz) 
and flat 
spectrum object (MG4 J200112+4352 in NED). Broad-band optical photometry of this source suggests 
variability also in
this waveband, while a spectroscopic follow-up observation provides the first source classification as
a BL Lac object. The source SED, as well as the overall characteristics and optical classification, 
point to a high frequency peaked blazar identification for 0FGL J2001.0+4352.
\end{abstract}

\begin{keywords}
X-rays: general,
Gamma-rays: general
X-rays: individuals: MG4 J200112+4352, 0FGL J2001.0+4352
\end{keywords}

\section{Introduction}

Following its launch in June 2008, the \emph{Fermi} gamma-ray telescope began to survey the sky at 
energies 
greater than 100 MeV. The catalogue of bright sources detected during the first 3 months of observations 
has recently been published by Abdo et al. (2009a) listing 205 objects with statistical significance 
greater than 10$\sigma$. Both Galactic and extragalactic populations are visible. The AGN class 
consisting of 121 sources is the largest in the catalogue. It contains mostly blazars (BL Lac, Flat 
Spectrum Radio Quasars and objects of uncertain classification), but also 2 radio galaxies (CenA and NGC 
1275) and the LMC. The Galactic population is primarily made of pulsars (29 members), plus 2 High 
Mass X-rays 
Binaries, a globular cluster (or most likely a source in it) and the Galactic centre. Thirty-seven of 
the \emph{Fermi} 
sources have no obvious counterparts at other wavelengths and are therefore still unclassified. 
Searching for counterparts of these sources is of course a primary objective of the survey work but it 
is made difficult by the good but still large \emph{Fermi} error circles. Cross-correlations with 
catalogues 
in 
other wavebands can be used as a useful tool with which to restrict the positional uncertainty of the 
objects detected by \emph{Fermi} and therefore to facilitate the identification process. In 
particular, observations at 
softer X-ray energies, where the positional accuracy is much better, could be an invaluable aid in the 
identification and classification process but the use of data at other wavebands, for example the 
radio, is also useful to identify peculiar and potentially/interesting candidate objects. Follow-up 
optical spectroscopy of likely candidates
can then provide classification of interesting sources and consequently a firm 
identification. In this Letter we 
report on the nature of the source 0FGL J2001.0+4352, one of the still unidentified objects in the 
\emph{Fermi} catalogue. We use 
unpublished \emph{XMM-Newton} and \emph{Swift}/XRT data to localise the X-ray 
counterpart of 
this high energy emitter and to characterise its X-ray spectrum. We further identify its radio 
counterpart and discuss its spectral properties. Finally, we report on an optical follow-up observation 
which provides the first source classification. 
All these data point to an identification of the MeV/GeV source with a blazar of BL Lac type.

\section{X-/gamma-ray data and results}

0FGL J2001.0+4352 is reported in the first \emph{Fermi} catalogue (Abdo et al. 2009a) as a source located 
close 
to the Galactic plane (7.1 degrees in Galactic latitude) and having RA(J2000)= 20$^{\rm h}$01$^{\rm
m}$05$^{\rm s}$.28 and Dec(J2000)= +43$^{\circ}$52$^{\prime}$15$^{\prime \prime}$.6 with a positional 
$95\%$ 
uncertainty of 0.069 degrees. Due to the faint flux and hard spectrum 
only an upper limit ($\leq 9.5 \times 10^{-8}$ ph cm$^{-2}$ s$^{-1}$) is reported in 
the 0.1--1 GeV band, while in the 1--100 GeV range the flux is 
(7.8$\pm$1.2) $\times 10^{-9}$ ph cm$^{-2}$ s$^{-1}$. The source is not associated with 
any previously known 
gamma-ray object. Using the method developed by 
Stephen et al. (2005,2006) to search for correlations between X-ray objects from various catalogues and 
\emph{Fermi} sources, we find a likely match between the XMM Slew source (Saxton et al. 2008) XMMSL1 
J200112.7+435255 and 0FGL J2001.0+4352. The XMM Slew source has a positional uncertainty of only 
4.51$^{\prime \prime}$; the source 0.2--12 keV flux is $\sim$$5.5\times10^{-12}$ erg cm$^{-2}$ s$^{-1}$ 
with $\sim$60$\%$ of the counts coming from the softest X-ray band (0.2--2 keV). 
On January 21, 2009 
\emph{Swift}/XRT (Gehrels et al. 2004) carried out an observation of 7.7 ks of the sky region containing 
the \emph{Fermi} source, thus allowing a comparison with the \emph{XMM-Newton} data.

\begin{figure} 
\centering 
\includegraphics[width=1.0\linewidth]{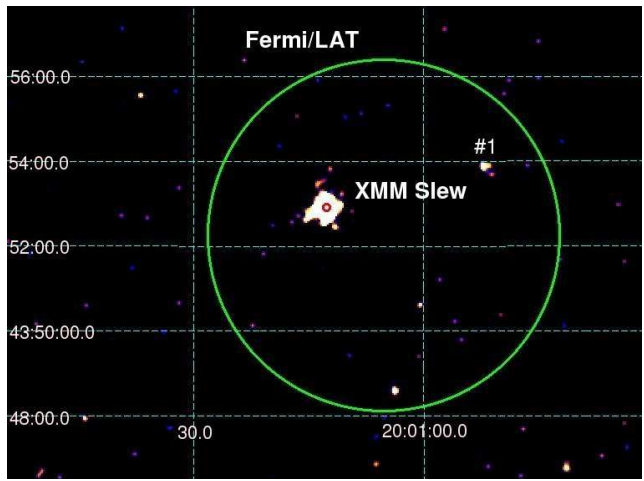}
\caption {\emph{Swift}/XRT 
0.3--10 keV image of 0FGL J2001.0+4352. The green and red circles correspond to the \emph{Fermi} and XMM 
Slew uncertainties. Source \#1 indicates the much weaker and softer object also detected by XRT.} 
\label{fig1} 
\end{figure}

XRT data reduction was performed using the XRTDAS standard data pipeline package ({\sc xrtpipeline} v. 
0.12.1), in order to produce screened event files. All data were extracted only in the Photon Counting 
(PC) mode (Hill et al. 2004), adopting the standard grade filtering (0--12 for PC) according to the XRT 
nomenclature. Images have been extracted in the 0.3--10 keV band and searched for significant excesses 
falling within or around the \emph{Fermi} error circle. A clear excess is observed in the XRT map with a 
significance of $\sim$18$\sigma$ at a position corresponding to RA(J2000)= 20$^{\rm h}$01$^{\rm 
m}$12$^{\rm s}$.88 and Dec(J2000)= +43$^{\circ}$52$^{\prime}$55$^{\prime \prime}$.0 and with an associated 
uncertainty of 3.7$^{\prime \prime}$ (90\% confidence level). This position is fully compatible with 
that reported in the XMM Slew catalogue; Figure~\ref{fig1} shows the XRT  0.3--10 keV image of the region 
surrounding 
0FGL J2001.0+4352 with superimposed both the \emph{Fermi} and \emph{XMM Slew} error circles. One more 
X-ray source (\#1 in the image) falls within the positional uncertainty of the MeV object but it is 
much weaker  
(the detection is around 3$\sigma$) and it disappears above 3 keV and it has no radio counterpart; it 
can reasonably be associated with a radio quiet AGN or a stellar object and not to a GeV emitter.

Events for spectral analysis were extracted within a circular region of radius 20$^{\prime \prime}$, 
centered on the source position, which encloses about 90\% of the PSF at 1.5 keV (see Moretti et al. 
2004). The background was taken from source-free regions close to the X-ray source of interest here, using 
circular regions with different radii in order to ensure an evenly sampled background. The source spectrum 
was then extracted from the corresponding event file using the {\sc XSELECT v.2.4} software and binned 
using {\sc grppha} in an appropriate way, so that the $\chi^{2}$ statistic could be applied. We used 
version v.011 of the response matrices and create the relative ancillary response file \textit{arf} using 
the task {\sc xrtmkarf v. 0.5.6}. The quality of the XRT data does not allow to fit more complex models 
than a simple power law filtered by the Galactic absorption which in the direction of the source is 
$3.91\times10^{21}$ cm$^{-2}$ (Kalberla et al. 2005; Dickey \& Lockman, 1990). This model provides a 
photon 
index $\Gamma=2.70\pm0.23$ and a 0.2--12 keV flux of $1.9\times10^{-12}$ erg cm$^{-2}$ s$^{-1}$, a factor 
of 2.9 lower than the one reported in the XMM Slew survey; we therefore conclude that the source is a 
persistent one but variable in flux over time.

\section{Radio counterpart}

\begin{figure} 
\centering 
\includegraphics[width=1.0\linewidth]{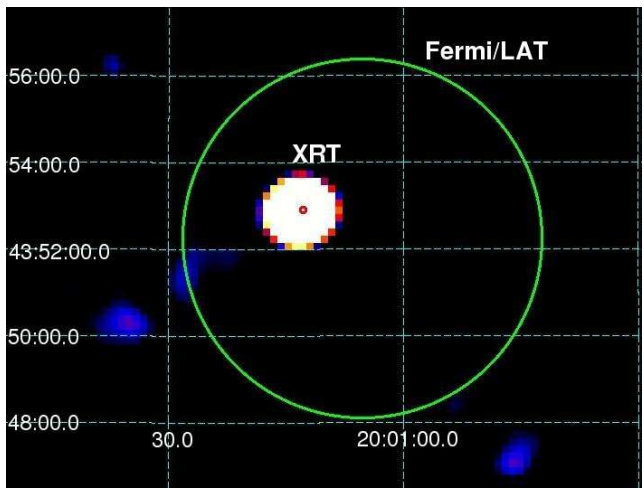} 
\caption{NVSS (20 cm) 
Image of the bright radio counterpart of the X-ray source possibly associated with 0FGL J2001.0+4352; 
the 
green and red circles correspond to the \emph{Fermi} and XRT positional uncertainties.  
The visible structure is likely due to background emission and is 
most probably unreal (see text).}  
\label{fig2}
\end{figure}

The radio image of the sky region of interest here has been taken from the NVSS (NRAO VLA Sky Survey, 
Condon et al. 1998) which covers the sky north of Declination (J2000.0) --40 degrees at 1.4 GHz. This 
survey provides 45 arcsecond FWHM angular resolution and nearly uniform sensitivity; the rms uncertainties 
in right ascension and declination are about 1 arcsecond for sources stronger than 15 mJy.  
Figure~\ref{fig2} is an NVSS image cut-out with overimposed the \emph{Fermi} error circle and XRT/XMM Slew 
positions. It is clear that a strong radio source is present in the \emph{Fermi} error circle and further 
coincides with the X-ray object detected both by \emph{Swift} and \emph{XMM-Newton}. This radio 
counterpart is located at RA(J2000)= 20$^{\rm h}$01$^{\rm m}$13$^{\rm s}$.15 and Dec(J2000)= 
+43$^{\circ}$52$^{\prime}$53$^{\prime \prime}$.1 (0.6$^{\prime \prime}$ uncertainty) and has a 1.4 GHz 
flux of 
$\sim$104 mJy. This source (named MG4 J200112+4352 in NED and hereafter) has been extensively observed at 
radio frequencies and appears in many radio catalogues although its nature has been poorly studied so far. 
To provide more information on this source, we have used SpecFind (Vollmer et al. 2005) and CATS (the 
on-line Astrophysical Catalogs support System, at http://cats.sao.ru/; see also Verkhodanov et al. 1997); 
both are tools used to cross-identify radio sources in various catalogues on the basis of self-consistent 
spectral index as well as position. This allows us to combine data at different frequencies (see 
Table~\ref{tab1}) 
and to estimate the source spectral index ($S(\nu)=A \nu^{\alpha}$), which in the case of MG4 J200112+4352 
is around 0.1 (see also NED); the source is therefore a flat spectrum object and thus possibly 
associated with a blazar type AGN.

\begin{table}
\begin{center}
\caption{Radio data.}
\label{tab1}
\begin{tabular}{cc}
\hline
\hline
Frequency & Flux \\
(MHz) & (mJy) \\
\hline
\hline
 325 &  0.223$\pm$0.011 \\
 1400 &  0.104$\pm$0.003 \\
 4830 &  0.147 \\
 4850 &  0.201--0.239 \\
 8400 &  0.215--0.223 \\
\hline
\hline
\end{tabular}
\end{center}
\end{table}

\begin{table}
\begin{center}
\caption{Optical/near infrared photometry.}
\label{tab2}
\begin{tabular}{cc}
\hline
\hline
Filter & Magnitude \\
\hline
\hline
 B &  15.62--17.89 \\
 V &  16.86--17.82 \\
 R &  14.12--16.03 \\
 I &  14.78--15.90 \\
J  &  15.02 \\
H  &  14.16 \\
K  &   13.48 \\
\hline
\hline
\end{tabular}
\end{center}
\end{table}

\section{Optical spectroscopy} 

\begin{figure}
\centering
\includegraphics[width=1.0\linewidth]{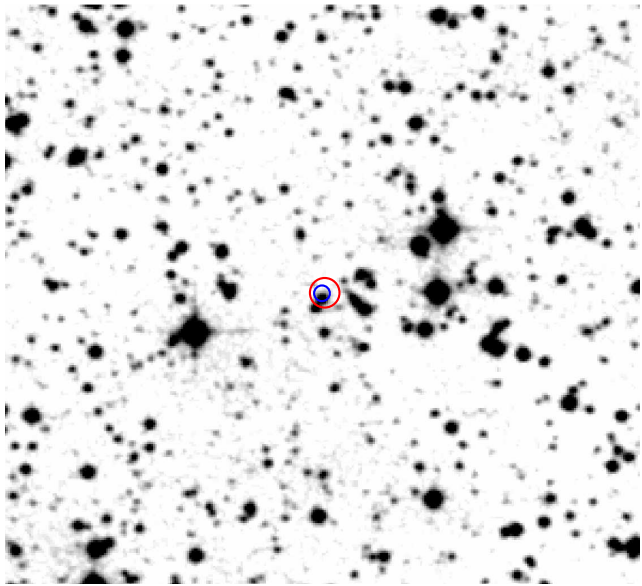}
\caption{DSS-II-Red image of the field of the gamma-ray source \emph{Fermi} 0FGL J2001.0+4352 with 
the 0.3--10 keV \emph{Swift}/XRT (smaller circle) and the 0.2--12
keV \emph{XMM-Newton} (larger circle) X-ray positions superimposed.
The only optical source positionally consistent with the XRT error circle, 
and with the NVSS radio source mentioned in the text, is the object
USNO--B1.0 1338--0359203 (at the centre of the image).
In the figure, North is at top and East is to the left. The field size is 
$\sim$5$'\times$5$'$.}  
\label{fig3}
\end{figure}

Within the respective X-ray/radio uncertainties, we find an 
optical/near-infrared source which has broad-band photometry available in the HEASARC database 
\footnote{{\tt http://heasarc.gsfc.nasa.gov/cgi-bin/W3Browse/w3browse.pl.}} and which we report in 
Table~\ref{tab2}. 
From this Table is evident the possibility that the source may be variable in the optical; even taking 
into 
account uncertainties due to differences in the observations set-up such as detector apertures, filter 
sensitivities, etc, the 
change in magnitude is sufficiently high to suggest a change in the source flux at least in the optical 
bands.
To firmly establish the nature of this X-ray/radio source, we next arranged and carried out a 
spectroscopic observation of its optical counterpart, USNO--B1.0 1338-0359203 (see Figure~\ref{fig3}). The 
source was spectroscopically observed on 14 March 2009 with the Bologna Astronomical Observatory 
1.52-metre ``G.D. Cassini'' telescope equipped with BFOSC, which uses a 1300$\times$1340 pixel EEV CCD. 
Observations started at 02:49 UT; three 1800 s spectra were acquired. In all exposures, Grism \#4 and a 
slit width of $2''$ were used, providing a 3500--8700 \AA~nominal spectral coverage. The use of this setup 
secured a final dispersion of 4.0~\AA/pix for all spectra.

After cosmic-ray rejection, the spectra were reduced, background 
subtracted and optimally extracted (Horne 1986) using 
IRAF\footnote{available at {\tt http://iraf.noao.edu/}}. Wavelength 
calibration was performed using He-Ar lamps acquired soon after each 
spectroscopic exposure; all spectra were then flux-calibrated using the 
spectrophotometric standard BD+28$^\circ$4211 (Oke 1990). Finally, the 
spectra were stacked together to increase the S/N ratio. The wavelength 
calibration uncertainty was $\sim$0.5~\AA~for all cases; this was checked 
using the positions of background night sky lines.

\begin{figure} 
\centering
\includegraphics[width=1.0\linewidth]{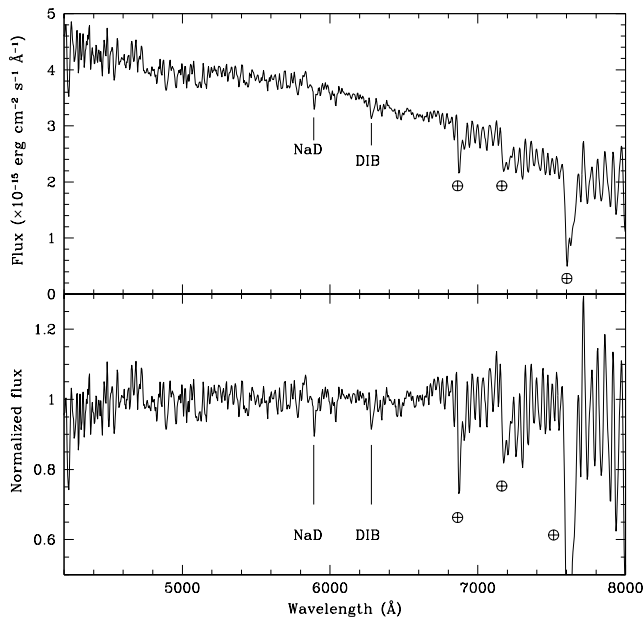} 
\caption{Spectrum of the optical counterpart of the \emph{Fermi} gamma-ray source
identified and discussed in this paper. The feature marked as NaD is the Galactic Na doubled 
at 5890  \AA. 
{\it Top panel}: spectrum
corrected for the intervening Galactic reddening assuming $E(B-V)$ =
0.562 mag (Schlegel et al. 1998) and smoothed using a Gaussian filter
with $\sigma$ = 3 \AA. The spectral features due to Galactic
interstellar absorption are labeled. The symbol $\oplus$ indicates
atmospheric telluric absorption bands.
{\it Bottom panel}: The same spectrum, normalized to the continuum
level.} 
\label{fig4}
\end{figure}

The optical spectrum of the  source (Figure~\ref{fig4}) shows a nearly 
featureless continuum, with only the atmospheric telluric lines and the 
Galactic diffuse interstellar band at 6280 \AA~apparent in absorption. No 
other line, either in absorption or in emission, is firmly detected. 
Regrettably this fact, also due to the relatively low 
signal-to-noise ratio of our data, does not allow us to pinpoint the 
redshift of this source. Nevertheless, at about 4800 \AA~there seems to be a slope change in the 
optical spectrum. This may be due to a non-thermal component which merges with the light from the host 
galaxy, almost completely diluting the Ca break at $\sim$4000 \AA.
In this case, we can give a rough estimate of the source redshift as $\sim$0.2.
After correction for the intervening 
Galactic reddening, $E(B-V)=0.562$ mag (Schlegel et al. 1998), the 
spectrum appears intrinsically blue and resembles those of BL Lac AGNs 
(see e.g. Sbarufatti et al. 2009 and references therein). This 
classification is also confirmed using the AGN classification 
approach of Laurent-Muehleisen et al. (1998).

\begin{figure} 
\centering
\includegraphics[width=1.0\linewidth]{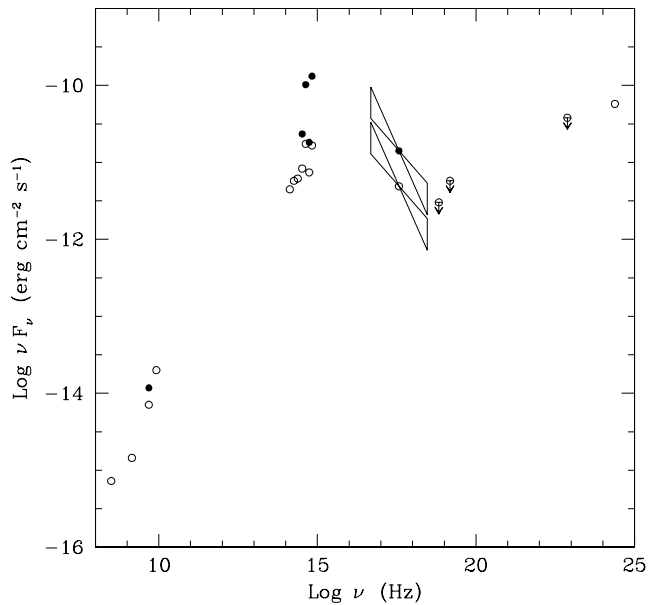} 
\caption{Broad-band non-simultaneous observational radio to gamma-ray SED of 0FGL J2001.0+4352, 
constructed with measurements presented in this paper (see text for details); open and filled circles 
represents the minimum and maximum value of the flux in each band. Optical, near-infrared and X-ray 
data are corrected for intervening Galactic absorption.} 
\label{fig5}
\end{figure}

\section{Is MG4 J200112+4352 the counterpart of 0FGL J2001.0+4352?}

MG4 J200112+4352 is clearly a bright radio/X-ray source optically identified here as a BL Lac object. 
This 
by itself provide strong evidence that it is the counterpart of the \emph{Fermi} source. The variability 
observed 
in the X-ray/optical band and the flat spectrum observed in radio further reinforce this conclusion. Are 
however the overall characteristics of the source compatible with other \emph{Fermi}
identified BL Lacs? Indeed in 
the gamma ($>$100 MeV) versus radio (8.4 GHz) flux diagram MG4 J200112+435 is located well in the region 
populated by BL Lac objects (see Fig. 14 in Abdo et al. 2009b). Also, the fluxes measured in the GeV 
bands 
suggest a hard source, i.e. more appropriate to a BL Lac object than to a flat spectrum radio quasar which 
is the other AGN typology detected by \emph{Fermi} (Abdo et al. 2009b). Finally, the location of the 
source in 
the X-ray (0.5--2 keV) versus radio (1.4 GHz) flux density plot indicates agreement with the positions of 
other \emph{Fermi} detected BL Lac objects and further suggests that MG4 J200112+4352 might be 
either a 
High Frequency Peaked Blazar (HBL) or an Intermediate Frequency Peaked Blazar (IBL). In the widely 
adopted scenario of blazars, a single population of high-energy electrons in a relativistic jet radiate 
from the radio/FIR to the UV- soft X-ray by the synchrotron process and at higher frequencies by inverse 
Compton scattering soft-target photons present either in the jet (Synchrotron Self-Compton [SSC] model), 
in the surrounding material (External Compton [EC] model), or in both (Ghisellini et al. 1998 and 
references therein). Therefore, a strong signature of the Blazar nature of a source is a double peaked 
structure in the Spectral Energy Distribution or SED, with the synchrotron component peaking anywhere from 
Infrared to X-rays and the inverse Compton extending up to GeV or even TeV gamma-rays. In 
Figure~\ref{fig5}, we construct the non-simultaneous SED of this enigmatic object by combining data 
collected in this work; to cover as many frequencies as possible we have also used upper limits obtained 
from the third \emph{INTEGRAL}/IBIS survey (Bird et al. 2007).
For the MeV/GeV data we have assumed a photon index of 2.0 as typically found in BL Lac objects
by \emph{Fermi}/LAT (Abdo et al. 2009b).
To deal with flux variability, we have plotted in 
the SED both minimum and maximum values found in the optical and X-rays wavebands. The source SED 
is clearly that of a HBL: the synchrotron emission is located close to the optical/UV band, 
the X-ray spectrum is steep (hence belonging to the synchrotron component) and the
Compton emission has a maximum in the GeV domain. We therefore conclude 
that all observational evidences point to the association of the \emph{Fermi} source 0FGL J2001.0+4352 
with the high frequency peaked BL Lac MG4 J200112+4352.

\section{Conclusions}

Through X-ray observations, we have been able to identify the counterpart of the newly discovered 
\emph{Fermi} 
source 0FGL J2001.0+4352 and to associate it with the radio source MG4 J200112+4352 so far optically 
unclassified. Optical follow-up observations have provided the first optical spectrum of the source, which 
is a new BL Lac object. The source has many peculiar features which are typical of a blazar: it is a radio 
bright object with a flat spectrum; it is likely variable at X-ray and optical frequencies; it is 
also most likely a HBL as suggested by its SED.

\section*{Acknowledgments}
We thank Roberto Gualandi for the Service Mode observations at the Loiano
telescope. We would like to thank the referee, Dr. P. Giommi, for his constructive comments and 
suggestions, which have improved the quality of this work.
This research has made use of data obtained from the High Energy Astrophysics Science Archive
Research Center (HEASARC), provided by NASA's Goddard Space Flight Center.
PP is supported by the ASI-INTEGRAL grant No. I/008/07.
We also acknowledge the use of public data from the \emph{Swift} data archive.

\end{document}